\documentclass[aps,preprint]{revtex4}
\usepackage{graphics,epsfig,amssymb}

\def\beq{\begin{equation}} 
\def\eeq{\end{equation}} 
\begin{document}

\title{Pairing correlations and eigenvalues of two-body density matrix in atomic nuclei}

\author{Michelangelo Sambataro$^a$ and Nicolae Sandulescu$^b$ \footnote{corresponding author, email: sandulescu@theory.nipne.ro}}
\affiliation{$^a$Istituto Nazionale di Fisica Nucleare - Sezione di Catania,
Via S. Sofia 64, I-95123 Catania, Italy \\
$^b$National Institute of Physics and Nuclear Engineering, P.O. Box MG-6, 
Magurele, Bucharest, Romania}

\begin{abstract}The quantum phases of superconductivity and superfluidity are characterised
 mathematically by the Yang concept of Off-Diagonal Long Range Order (ODLRO), related to the
existence of a large eigenvalue of the density matrix. We analyse how the Yang
criterion applies for various Hamiltonians commonly employed to describe superfluid-type 
correlations in atomic nuclei. For like-particle pairing Hamiltonians the behaviour of the
largest eigenvalue of two-body density matrix shows a clear evidence for a transition 
between a normal to a paired phase. However, this is not the case for the isoscalar 
proton-neutron pairing interactions acting in self-conjugate nuclei.

\vskip 0.4cm

Keywords: Pairing in nuclei, density matrix, Yang criterion.

\end{abstract}

\maketitle

\section{Introduction}
 
In infinite many-body systems the long-range correlations associated with superconductivity and
superfluidity are connected  to the properties of the eigenvalues of density matrices.
More specifically, according to the criterion proposed  by Penrose and  Onsager \cite{penrose, penrose_onsager}, 
 a system of interacting bosons is in the  Bose-Einstein (B.E.) condensation phase whenever
the reduced one-body density matrix has one large eigenvalue of the order of $N$, where $N$ is the particle number. 
The largest eigenvalue is associated with the number  of ``condensed" bosons and the corresponding eigenfunction
represents the "wave function of the condensate".  
This criterion generalizes
the common definition of   B.E. condensation of  non-interacting boson gases, in which the bosons are condensed
in the lowest single-particle level. The criterion of Penrose and Onsager was
later on extended by C. N. Yang \cite{yang} to more general many-body systems. 
In particular, the phenomena of superconductivity and superfluidity in fermionic systems are characterized by the
existence of a large  eigenvalue of the reduced two-body density matrix. In this case, the largest eigenvalue
is related to the ``condensed" Cooper pairs. In spatially extended systems the existence of a large eigenvalue 
in the reduced density matrix implies, in  coordinate representation,  the existence  of  
off-diagonal long-range order (ODLRO) \cite{yang}.  The concept of ODLRO is of fundamental 
importance because it characterizes in what sense the superfluidity and superconductivity are similar
phenomena. In particular, as shown by C. N. Yang, the ODLRO  gives rise to the quantized magnetic flux
and predicts its required value for the classical superconductors.
 
In atomic nuclei the long range correlations of superfluid type are commonly treated in the 
framework of BCS theory \cite{bcs}. 
BCS-like models are able to describe in a simple way many fundamental properties of nuclei, such as moment 
of inertia, odd-even mass differences, excitation spectra,  pair transfer reactions, etc. \cite{bohr_mottelson}. 
However, since  BCS is a theory designed for infinite systems, it has a limited validity when
applied to finite quantum systems such as atomic nuclei. Mathematically, this can be seen 
from the fact that the BCS ansatz  is an exact solution of the standard pairing Hamiltonian 
only in the thermodynamical limit, which is quite far from the properties 
of atomic nuclei (e.g., see \cite{richardson}). Phenomenologically, the most evident drawback 
of BCS, when applied to nuclei, is its prediction of a type II phase transition from the normal to 
the superfluid phase, which is not expected to appear in  finite size systems. 
To avoid these drawbacks, the pairing problem is usually treated in particle-number 
conserving approaches, such as particle-number projected BCS (PBCS) \cite{bayman,pbcs} 
or  shell model-like models.  However, going beyond BCS comes with one  difficulty: how
to identify in the structure of the wave functions the presence of pairing correlations of superfluid
type.  One possible way is to use the Yang criterion based on the eigenvalues of reduced two-body density matrix. 
How this criterion works in nuclei is not well documented. To our knowledge, the only application of 
Yang criterion to nuclei, which provides details on the spectral decomposition of the density matrix, 
is in relation to finite-temperature shell model calculations \cite{langanke}. 
The scope of this paper is to analyse the properties of the eigenvalues of two-body density matrix for the ground
state (i.e. at zero temperature) of various pairing Hamiltonians, commonly used to describe pairing in nuclei, 
and to assess the ability of Yang criterion for identifying the onset of pairing correlations in finite systems. 
In the first part of the paper we shall discuss, from the perspective of Yang criterion, the case of like-particle pairing and then we shall focus on the proton-neutron pairing correlations in self-conjugate nuclei, which is presently one of the most debated issues in nuclear structure \cite{frauendorf,sagawa}.

\section{Density matrix and pairing: general formalism}

\subsection{Density matrix in  coordinate representation}

For the sake of completeness, we start by presenting  how the two-body
density matrix is usually employed to describe pairing correlations in fermionic
systems \cite{leggett}.
The reduced two-body density for a system of N fermions is given by the expression
\beq
\rho^{(2)} (r_1, r_2 ; r'_1,  r'_2 ) =
\langle \Phi^{(N)}_0  | \hat{\Psi}^+ (r_1) \hat{\Psi}^+ (r_2) 
\hat{\Psi}  (r'_2)
\hat{\Psi} (r'_1 ) | \Phi^{(N)}_0 \rangle ,
\eeq
where $| \Phi^{(N)}_0\rangle$ is the ground state of system, $\hat{\Psi}^+ (r)$ is the 
fermonic creation operator while $r={\vec{r},\sigma}$ denotes the spatial and spin variables. 

The two-body density
can be diagonalized and expressed in terms of its eigenvalues and eigenvectors as
\beq 
\rho^{(2)} (r_1 , r_2 ; r'_1 ,  r'_2 ) =
\sum_n \lambda_n \phi^*_n(r_1 , r_2 ) \phi_n(r'_1 , r'_2 ) ,
\eeq
where the functions $\phi_n(r_1 , r_2 )$ are mutually orthogonal and 
normalized. The definition of the reduced two-body density imposes the following constraint on the eigenvalues
\beq
\sum_n \lambda_n= N(N-1) ,
\eeq
where N is the number of particles. 

Eq. (3) is by itself compatible with the occurrence of one or more  eigenvalues of the order of $N^2$. However, due to the Fermi statistics, the largest possible
eigenvalue turns out to be of the order of N. More precisely, as shown by Yang \cite{yang}, 
for a system of N fermions distributed on M single-particle states the maximum 
possible value for the eigenvalues is 
\beq
\lambda_{max}=\frac{N(M-N+2)}{M} .
\eeq

According to the Yang criterion, a Fermi system is in a Cooper pairing phase if among the
eigenvalues of $\rho^{(2)}$ there is one eigenvalue of the order of N and the others 
are of the order of unity. If there is more than one eigenvalue of the order of N, then 
the condensate is called ``fragmented".  What really means ``of the order of N" in the case
of finite systems such as atomic nuclei is one of the question that will be addressed in this study.
 
With the largest eigenvalue $\lambda_0$  and the corresponding eigenvector 
$\phi_0(r_1 , r_2 )$ one defines the order parameter or the ``wave function of the condensate"
\beq
F(r_1, r_2 ) = \sqrt{\lambda_0} \phi_0(r_1 , r_2 ).
\eeq
By integrating $|F|^2$ over the space variables and performing the sum over the spin projections one gets the so-called
number of condensed pairs.  Since the function $\phi_0 (r_1 , r_2 )$ is normalized, the number of
condensed pairs is actually equal to $\lambda_0$.  

When one eigenvalue is much larger than the others, the two-body density has 
special localisation properties. This can be seen by considering in Eq. (2) the limit of large 
separation  between $(\vec{r}_1+\vec{r}_2)/2$ and  $(\vec{r'}_1+\vec{r'}_s)/2$ and keeping
finite values for $| \vec{r}_1-\vec{r}_2 |$ and $| \vec{r'}_1-\vec{r'}_2 |$. In this limit,
from the r.h.s of Eq. (2) is surviving only the term corresponding to the largest
eigenvalue because the contribution of the other eigenvalues is expected to vanish
due to the destructive interference \cite{leggett}.  Therefore, in this limit, which can
be fulfilled if the system is extended, the two-body density reduces to
\beq
\rho^{(2)} (r_1, r_2 ; r'_1 ,  r'_2 ) 
= F^*(r_1 , r_2 ) F(r'_1 , r'_2 )  .
\eeq
This equation implies that in the presence of a large eigenvalue, as in  
the Cooper pairing  phase,  the off-diagonal matrix elements of the density matrix in the 
coordinate representation remain finite for large distance between the center of mass of the 
two  sets of variables . This is the concept of  off-diagonal long -range order (ODLRO) introduced by Yang \cite{yang}. Since for extended systems the existence of the largest eigenvalue is equivalent to the existence of ODLRO, the latter  signals the Cooper pairing phase.

For the $s$-wave pairing the relevant correlations are the ones between  the fermions with spin-up and
spin-down, i.e., for  $\sigma_1 =-\sigma_2=1/2$. The corresponding order parameter (5) is commonly 
written as $F(\vec{R},\vec{r}; \sigma_1=-\sigma_2=1/2)$, where  $\vec{R}=(\vec{r}_1+\vec{r}_2)/2$ 
and $\vec{r}=\vec{r}_1-\vec{r}_2$ are the center of mass and relative coordinates, respectively. In particular, 
the function $F(\vec{R},\vec{r}=0)$ is proportional to the Ginzburg-Landau order parameter \cite{ginzburg_landau}.  
In the BCS-like approximation the function $F$ is the so-called pairing tensor in the coordinate
space and it is defined by $F^{BCS} (r_1, r_2) = \langle BCS | \hat{\Psi}  (r_1)
\hat{\Psi} (r_2 ) | BCS \rangle$. In finite systems such as atomic nuclei the spacial properties 
of  $F(\vec{R},\vec{r})$ are rather complex because they are influenced not only by the pairing 
interaction but also by the finite size of the system (e.g., see Refs \cite{ns_kappa0,ns_kappa}).

\subsection{Density matrix in the configuration representation}

The  two-body density matrix in coordinate representation can be reformulated in terms of the
matrix elements of two-body density in configuration representation. This can be achieved
by expanding in Eq.(1) the particle operators in a single-particle basis. Hence, using the
expansion $\hat{\Psi}^+(r)= \sum_i \varphi^*_i (r) \hat{c}^+_i$ the reduced two-body
density can be written as
\beq
\rho^{(2)} (r_1, r_2; r'_1,  r'_2 ) =
\sum_{i_1, i_2, i'_1, i'_2}
\varphi^*_{i_1} (r_1) \varphi^*_{i_2} (r_2)
\varphi_{i'_1} (r'_1) \varphi_{i'_2} (r'_2)
 \rho^{(2)}_{i_1, i_2; i'_1, i'_2} ,
\eeq
where
\beq
\rho^{(2)}_{i_1, i_2; i'_1, i'_2} =
\langle \Phi^{(N)}_0  | \hat{c}^+_{i_1}  \hat{c}^+_{i_2} 
\hat{c}_{i'_2} \hat{c}_{i'_1} | \Phi^{(N)}_0  \rangle 
\eeq
are the matrix elements of the reduced two-body density in the configuration representation.

The eigenvalues and the eigenvectors of this matrix are defined by
\beq
\sum_{i'_1,i'_2} \rho^{(2)}_{i_1, i_2; i'_1, i'_2} f^{(n)}_{i'_1,i'_2} = \lambda_n f^{(n)}_{i_1,i_2}. 
\eeq 
With these eigenvalues and eigenvectors one can then write the spectral decomposition of
the density matrix in the configuration representation
as
\beq
\rho^{(2)}_{i_1, i_2; i'_1, i'_2} = \sum_n \lambda_n f^{(n)*}_{i_1,i_2} f^{((n)}_{i'_1,i'_2}
\eeq
This decomposition is analogous of the decomposition (2) in the coordinate representation.
The relation between them is obtained by replacing (10) into Eq.(7). One thus see that 
\beq
\phi_n(r_1, r_2 ) \equiv \sum_{i_1,i_2} f^{(n)}_{i_1,i_2} 
\varphi_{i_1} (r_1) \varphi_{i_2} (r_2) .
\eeq
The mutual orthogonality and the normalization of the functions $\phi_n$
is assured by the orthogonality and the normalization of the single-particle basis states $\varphi_i$
and by the properties of the eigenvectors $f^{(n)}_{i_1,i_2}$. 
As a consequence, in the expansions (2) and (9) the eigenvalues are the same, as expected.

\subsection{Density matrix and two-particle transfer}

From Eq. (10) it can be seen that the eigenvalues of the density matrix can be written as
\beq
\lambda_n = \langle \Phi^{(N)}_0 | \tilde{P}^+_n \tilde{P}_n | \Phi^{(N)}_0 \rangle
\eeq
where
\beq
\tilde{P}^+_n = \sum_{i_1,i_2} f^{*(n)}_{i_1,i_2} \hat{c}^+_{i_1}  \hat{c}^+_{i_2} 
\eeq
is a collective pair operator built with the amplitudes $f^{(n)}_{i_1,i_2}$ 
corresponding to the eigenvalue $\lambda_n$. 
Making use of the completeness of the eigenfunctions for the system with $(N-1)$, i.e., 
\beq
\hat{1}= \sum_\nu | \Phi^{(N-1)}_\nu \rangle \langle \Phi^{(N-1)}_\nu |
\eeq
the Eq. (12) can be written as
\beq
\lambda_n = \sum_{\nu} |\langle \Phi^{(N-1)}_\nu| \tilde{P}_n | \Phi^{(N)}_0 \rangle |^2.
\eeq
According to the expression above, the eigenvalue $\lambda_n$ is the two particles transfer intensity associated with $\tilde{P}_n$. This is
defined as the sum of the squares of the transfer amplitudes between the ground state of the 
system with N pairs and all the possible eigenstates of the system with (N-1) pairs which can be 
reached via the collective pair operator $\tilde{P}_n$. It can be proved that  the collective
operators  $\tilde{P}^+_n$ are the ones which, among all possible pair 
operators, maximize the transfer intensity defined above.

From the relation (15) one sees that the maximum transfer intensity is associated with the
collective pair operator $\tilde{P}^+_0$  built with the amplitudes $f^{(0)}_{i_1,i_2}$
of the maximum eigenvalue $\lambda_{0}$. This pair operator has the highest collectivity 
because the amplitudes $f^{(0)}_{i_1,i_2}$ are expected to act coherently (i.e. to have the same sign)
when the system has long-range correlations, as in the case of BCS-like pairing phase.

The eigenvalues of the density matrix can also be related to the two-particle transfer amplitudes
associated with the non-collective pair operators. Indeed, by taking into account that the sum of the eigenvalues is equal to the trace
of the two-body density  matrix and using the completeness  relation (14)  one gets:
\beq
\sum_n \lambda_n =  \sum_\nu \sum_{i_1,i_2} | \langle \Phi^{(N-1)}_\nu
|\hat{c}_{i_1}  \hat{c}_{i_2} | \Phi^{(N)}_0 \rangle |^2.
\eeq

Summarizing, the largest eigenvalue $\lambda_0$ of the two-body density matrix
corresponds to the maximum transfer intensity between the ground state of the system with N
pairs and all the states of the system with N-1 pairs. This maximum transfer intensity is generated by the most coherent pair
transfer operator built with the eigenfunctions of the two-body density matrix. We have also seen that the sum of the eigenvalues $\lambda_n$ can be related to the two particle transfer amplitudes associated
with the non-collective pair operators. These amplitudes are the ones associated with the physical
two-particle transfer process.

\section{ Eigenvalues of density matrix for like-particle pairing interactions}

When the pairing is among like-particle, such as electrons in superconducting grains and neutrons or protons in atomic nuclei,
of special interest is the density matrix corresponding to time-reversed states, defined by
\beq
\rho^{(2)}_{i,i'} =
\langle \Phi^{(N)}_0 | \hat{c}^+_{i} \hat{c}^+_{\bar{i}}
\hat{c}_{\bar{i'}} \hat{c}_{i'} | \Phi^{(N)}_0 \rangle ,
\eeq
where $\bar{i}$ denotes the time-reversed of the state $i$. For spherically-symmetric states,
labeled by the standard quantum numbers $i=\{ n,l,j; m \} \equiv \{a; m \} $, one
usually defines
the density matrix corresponding to J=0 pairs
\beq
\rho^{(2)}_ {a,a'}=
\langle \Phi^{(N)}_0  | P^+_a P_{a'}  | \Phi^{(N)}_0  \rangle ,
\eeq
where 
\beq
P^+_a = \frac{1}{\sqrt{2}}[(\hat{c}^+_{a} \hat{c}^+_{a}]^{J=0}.
\eeq

The pairing correlations among like-particle fermions are commonly described by the Hamiltonian
\begin{equation}
H=\sum_i  \epsilon_i N_i - \frac{1}{2} \sum_{a, b} V(a,b) \sqrt{(2j_a+1)(2j_b+1)} P^+_a P_b .
\end{equation}
In the first term, $\epsilon_i$ and $N_i$ are the energy and the particle number operator
relative to the single-particle state $i$, respectively. The pairing interaction is expressed in terms of the pair operators
 introduced above.
In the following we will analyze the eigenvalues of the two-body density matrix corresponding to the ground state of the
Hamiltonian (20) calculated for various single-particle states and pairing interactions. 

We start by considering the  SU(2) seniority model, i.e. a system of N fermions sitting on
a single level with angular momentum $j$ and energy $\epsilon_j =0$ and interacting through a constant strength
pairing interaction, $V(a,b)=g$. In this case the Hamiltonian (20) can be solved exactly,
the ground state for even $N$ being simply proportional to $(P^+_j)^{N/2}|0\rangle$, and  the 
corresponding two-body density matrix is
\beq
\rho^{(2)}_{sen}=N(M-N)/2M+N/M ,
\eeq
where $M=2j+1$ is the degeneracy of the level. The last term in the equation above represents
the mean field contribution to the density matrix. This can be seen from the fact that this
term survives when the level is completely filled, i.e., $M=N$, when there are no genuine
pairing correlations. 
Since in the seniority model the particles are subject to the maximum
pairing correlations, the expression (21) is equal to the maximum eigenvalue (4) of the density matrix, apart from a factor $1/2$.  This factor comes from the fact that in the density matrix (18) it is used a factor $1/\sqrt{2}$ in front of the pair operator (19).
 
We discuss now the two-body density matrix in the BCS approximation. Its general expression is
 \beq 
\rho^{(2)}_{a,a'} = \langle BCS | P^+_a P_{a'}| BCS \rangle>
=  \hat{\kappa}_a \hat{\kappa}_{a'}+\delta(a,a') v_a^4 .   
\eeq  
In the equation above  $\hat{\kappa}_a = \sqrt{(2j_a+1)/2}\kappa_a$,
where  $\kappa_a= u_a v_a$ is the diagonal part of the pairing tensor, 
$v_a^2$ is the occupation probability of the state $a$ of angular momentum $j_a$ and  $u_a^2=1-v_a^2$. 
 
The last term in (22) is the mean field contribution. Without this term the two-body density reduces to
\beq
\tilde{\rho}^{(2)}_{a,a'}= \hat{\kappa}_a \hat{\kappa_{a'}}
\eeq
It can be easily shown that $\tilde{\rho}^{(2)}$ has only one non-zero eigenvalue
equal to 
\beq
\tilde{\lambda}_0 = \frac{1}{2} \sum_a (2j_a +1) u^2_a v^2_a .
\eeq
 The quantity $\tilde{\lambda}_0$, without the factor $1/2$ in front, is sometimes called the ``number of condensed pairs".
As expected, in the limit of vanishing pairing correlations, when $v^2_a$ are equal to
one or zero, $\tilde{\lambda}_0=0$.

 In the case of N particles distributed over the M degenerate states of a single-j orbit, when $v^2=N/M$,  the two-body density 
matrix in the BCS approximation becomes
\beq
\rho^{(2)}_{BCS}=N(M-N)/2M+N^2/M^2 .
\eeq
The first term corresponds to $\tilde{\lambda}_0$ while the last term is the mean field contribution.
In the mean field limit, i.e. N=M, the two-body density is equal to 1, as in the case
of the exact expression (21).

We shall now analyze the properties of two-body density matrix for two non-trivial 
pairing Hamiltonians, comparing the results obtained in various approximations.
First we discuss a system formed by 12 particles distributed over 12  doubly degenerate levels with 
equidistant energies and interacting among them with a state-independent interaction. This system is
described by the Hamiltonian (20) by taking $\epsilon_i=id$, with $i=1,2,..12$ and $d$ being the distance between the levels, $j_i=1/2$ and $V(a,b)=g$.
This Hamiltonian gives a schematic description of axially deformed nuclei and also of superconducting
metalic grains \cite{schechter,delft}.
 
\begin{figure}[ht]
\centering
\begin{tabular}{cc}
\epsfig{file=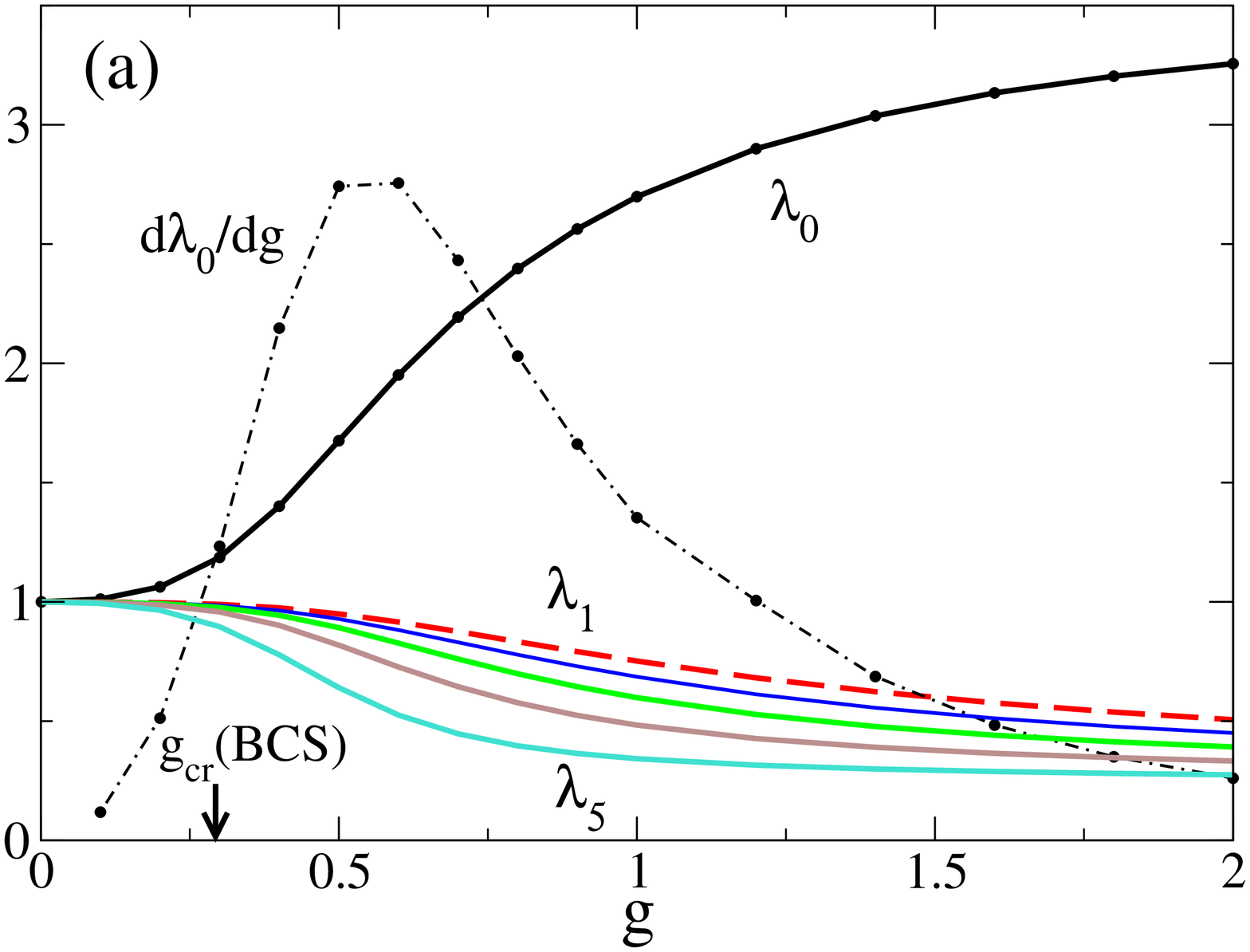,width=2.5in,height=3.5in,angle=-90} &
\epsfig{file=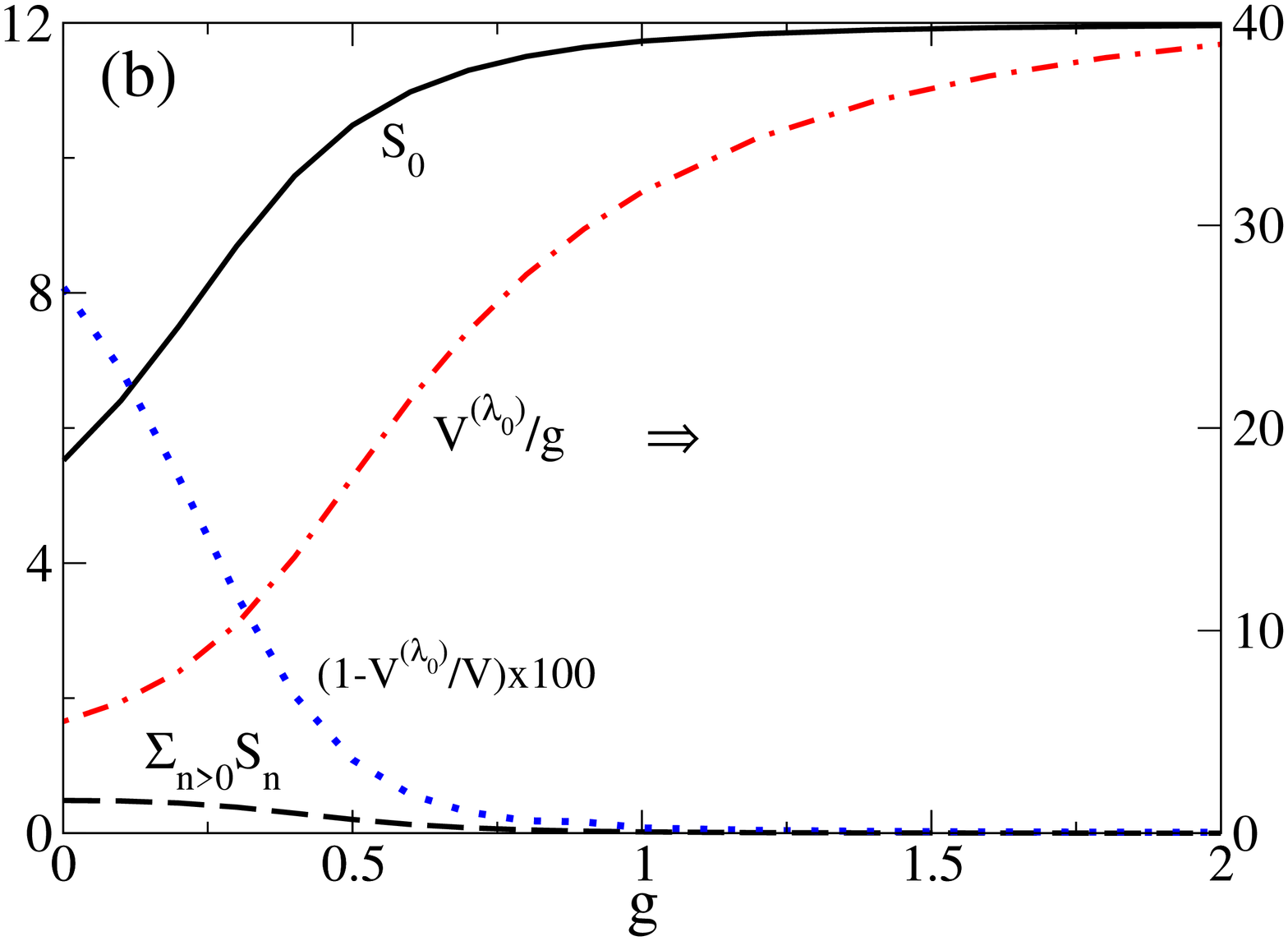,width=2.5in,height=3.5in,angle=-90} \\
\end{tabular}
\caption{Exact results for a system of 12 particles moving in 12 doubly degenerate levels.
(a) The largest 6 eigenvalues of the two-body density matrix as a function of the
pairing strength; for the largest eigenvalue $\lambda_0$ it is also shown its derivative to the pairing
strength. (b) Various quantities (see text) related to the contribution of the largest eigenvalue to
the average of the pairing interaction.} 
\end{figure}

  For the system described above the eigenvalues of the two-body density matrix can be calculated
  exactly by solving the Richardson equations \cite{richardson}. 
  In Fig. 1a are shown, as a function of pairing strength,  the largest 6 eigenvalues of the density 
  matrix (18). In the non-interacting
 limit, i.e.  $g=0$,  these eigenvalues are equal to 1 and correspond to the 6 occupied levels.  
 When the strength 
  of the pairing interaction is switched on, one eigenvalue becomes greater than 1 while the others smaller
  than 1. It is also observed that, at variance with the infinite systems, the second  largest eigenvalue 
  $\lambda_1$ is not too much different from the largest eigenvalue $\lambda_0$.

The dependence of $\lambda_0$ on the pairing strength indicates how the system evolves from 
a normal Hartree-Fock (HF)- like state to a paired phase. In order to better visualize 
the evolution of the 
largest eigenvalue with $g$, in Fig. 1a we display its derivative with respect to the strength. 
One can observe that the system evolves quite fast to a paired phase, the fastest variation taking
place for $g > g_{cr}$, where $g_{cr}$ is the critical value of the strength in the BCS approximation. 
It is interesting to notice that $d\lambda_0/dg$ is quite similar in  shape to the invariant correlation 
entropy,  which is another way of analyzing the evolution of a system toward a paired  phase \cite{volya}. 

With the two-body density matrix one can evaluate the expectation value of any two-body operator.
Of special interest is the expectation value of the two-body interaction in the ground state of
the system
\beq
V \equiv \langle \Phi_0  | \hat{V} | \Phi_0 \rangle = \sum_{ik} V_{ik} \rho_{ik} ,
\eeq
which can be written as
\beq
V  = \sum_n \lambda_n \sum_{ik} V_{ik} f^{(n)}_i f^{(n)}_k 
                     \equiv \sum_n V^{(\lambda_n)} .
\eeq

In the case of a state independent pairing interaction, when  $V_{ik}= g \sqrt{(2j_i+1)(2j_k+1)}$, one gets
\beq
V = g \sum_n \lambda_n (\sum_{i} \sqrt{2j_i+1} f^{(n)}_i)^2 .
\eeq
For the system analyzed here all the states have j=1/2 and therefore for each eigenvalue $\lambda_n$ the
important quantity is $S_n=(\sum_i f^{(n)}_i)^2$. In Fig. 1b it is shown the quantity 
$V^{(\lambda_0)}/g = \lambda_0 S_0$, which represents the contribution of the largest eigenvalue
to the average of the pairing interaction in units of the pairing strength, and the quantity
$(1-V^{(\lambda_0)}/V) \times 100$. From the latter one can notice  that the main contribution 
to $V$ comes from the largest eigenvalue, in spite of the fact that the other eigenvalues
are also quite large, especially in the weak coupling regime.  This is due to the fact that the components $f^{(n)}_i$  relative to  $\lambda_n$ act coherently only for the
largest eigenvalue. More precisely, all the components $f^{(0)}_i$ turn out to have the same sign, 
which is not the case for the eigenfunctions corresponding to the other eigenvectors. Due to this fact, 
the quantity $S_0$ is much larger than $S_{n>0}$. This can be seen in Fig. 1b where it is plotted $S_0$ and $\sum_{0<n \le 5} S_n$. 

\begin{figure}[ht]
\centering
\begin{tabular}{cc}
\epsfig{file=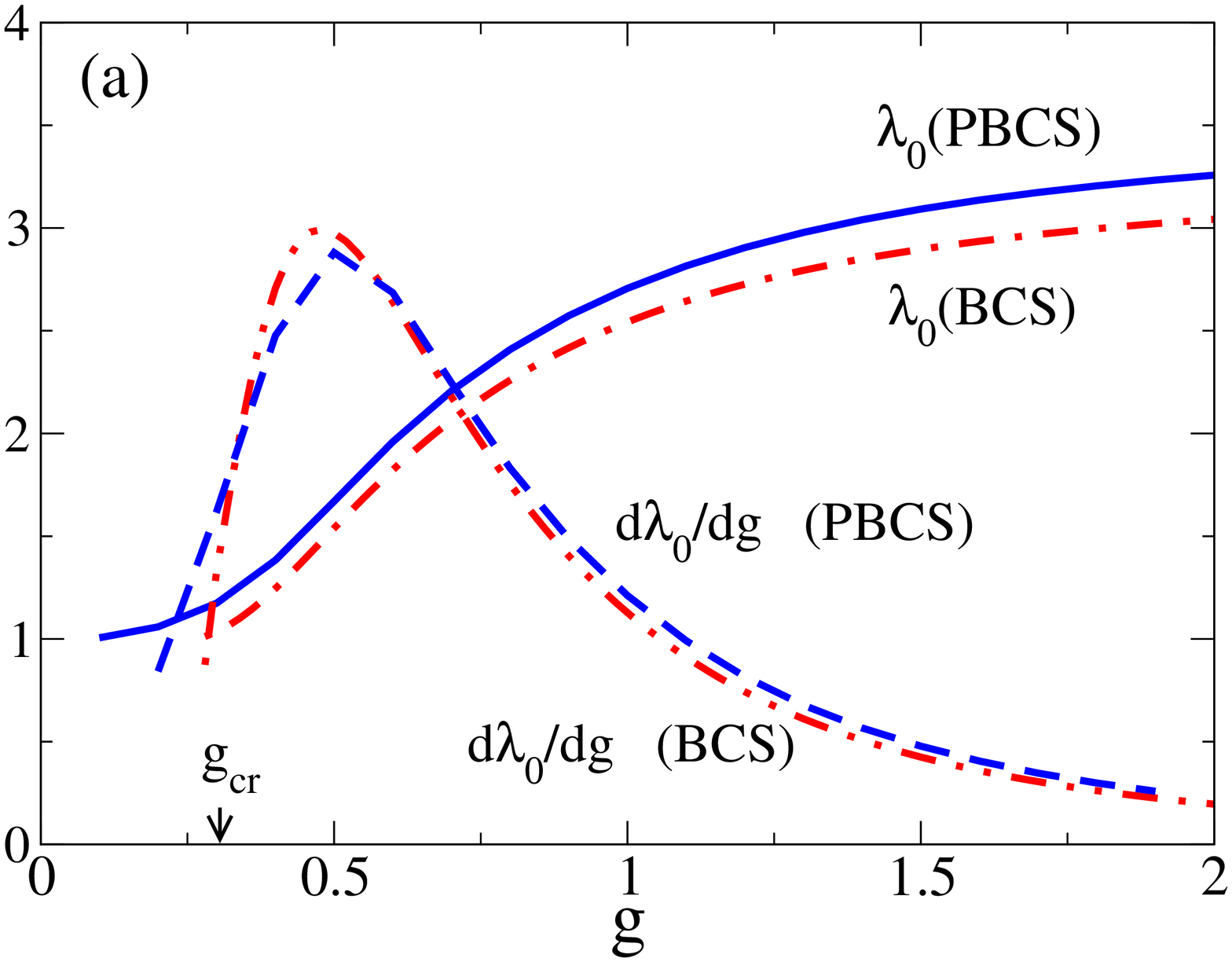,width=2.5in,height=3.5in,angle=-90} &
\epsfig{file=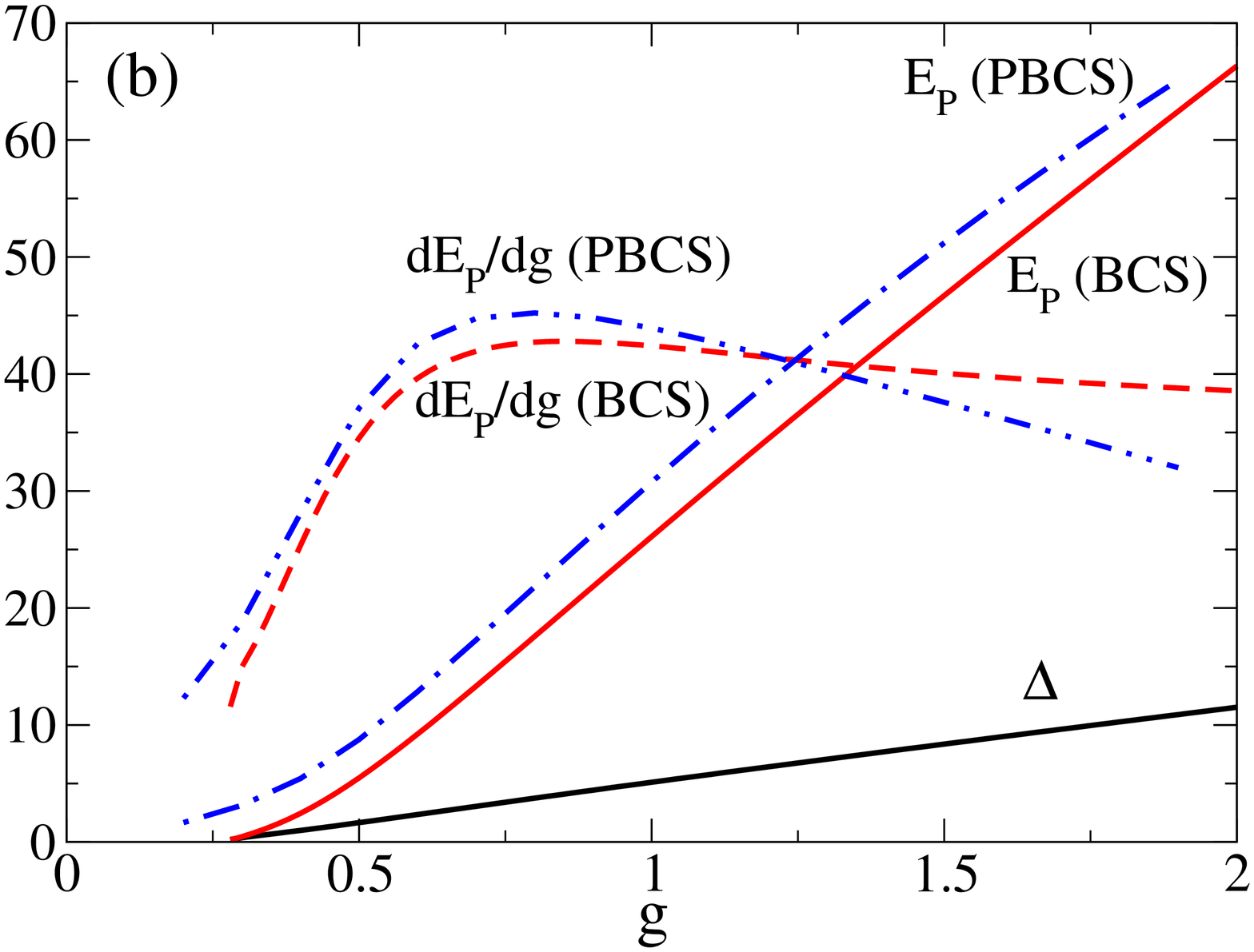,width=2.5in,height=3.5in,angle=-90} \\
\end{tabular}
\caption{
BCS and PBCS results for a system of 12 particles moving in 12 doubly degenerate levels. 
(a) The largest eigenvalue of the two-body density matrix and its derivative to
the pairing strength. (b) The pairing energies and their derivatives to the pairing strength. 
$\Delta$ is the BCS pairing gap.}
\end{figure}

In Fig. 2a is shown, for the same system, the largest eigenvalue calculated in the BCS 
approximation.  In BCS there is a phase transition from the normal phase to the BCS pairing
phase at $g=g_{cr}$. Then, for  $g>g_{cr}$ the largest eigenvalue is increasing rather fast, in a 
similar manner as for the exact solution. In particular, $d\lambda_0/dg$ is reaching the
maximum value in the same region of the strength as in the case of the exact solution.

In Fig. 2a are also shown the results obtained in the particle-number projected BCS (PBCS) approximation
\cite{bayman,pbcs}. 
Within PBCS the ground state of the system is described by the ansatz
\beq
| PBCS \rangle = (\Gamma^+)^{N/2} | 0 \rangle ,
\eeq
where $\Gamma^+ = \sum_i x_i a^+_i a^+_{\bar{i}}$ is the collective Cooper pair.  The variational parameters 
$x_i$ are determined from the minimization of the average of the Hamiltonian in the PBCS state, calculated 
with the method of recurrence relations \cite{sandulescu_errea}. Since the PBCS is built by applying the same 
pair operator on $| 0 \rangle $, the PBCS state is usually called a pair condensate. When the pairing correlations
are vanishing, the PBCS state becomes a HF-like state. 
Comparing Fig. 2a  with Fig. 1a one can observe  that  the  PBCS results follow closely the exact results
in all the coupling regimes, from weak to strong coupling. One can also notice  that, as in the case of the
exact solution, the PBCS results evolve smoothly across the region $g=g_{cr}$ where BCS  predicts a phase 
transition. 

\begin{figure}[ht]
\centering
\begin{tabular}{cc}
\epsfig{file=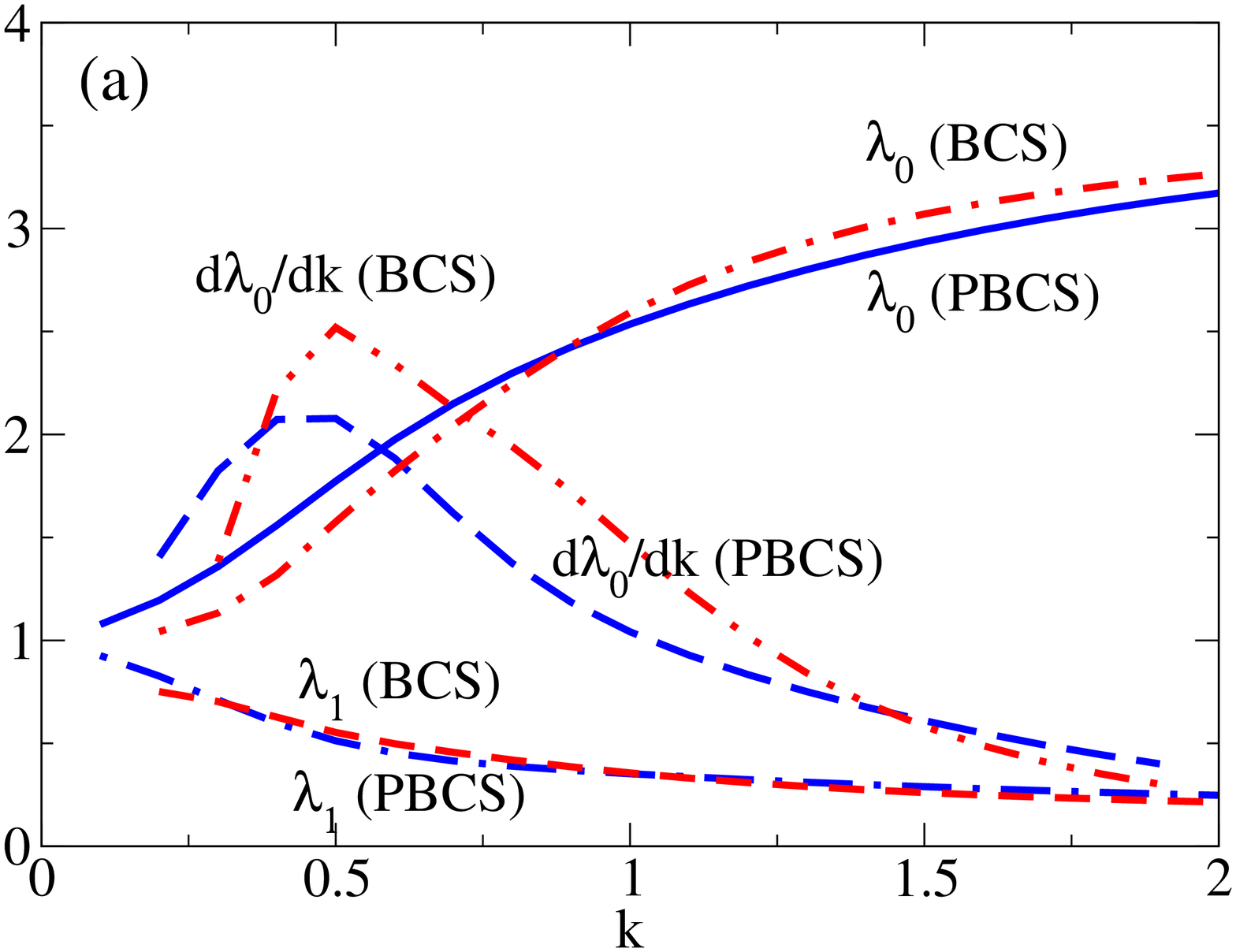,width=3in,height=3.5in,angle=-90} &
\epsfig{file=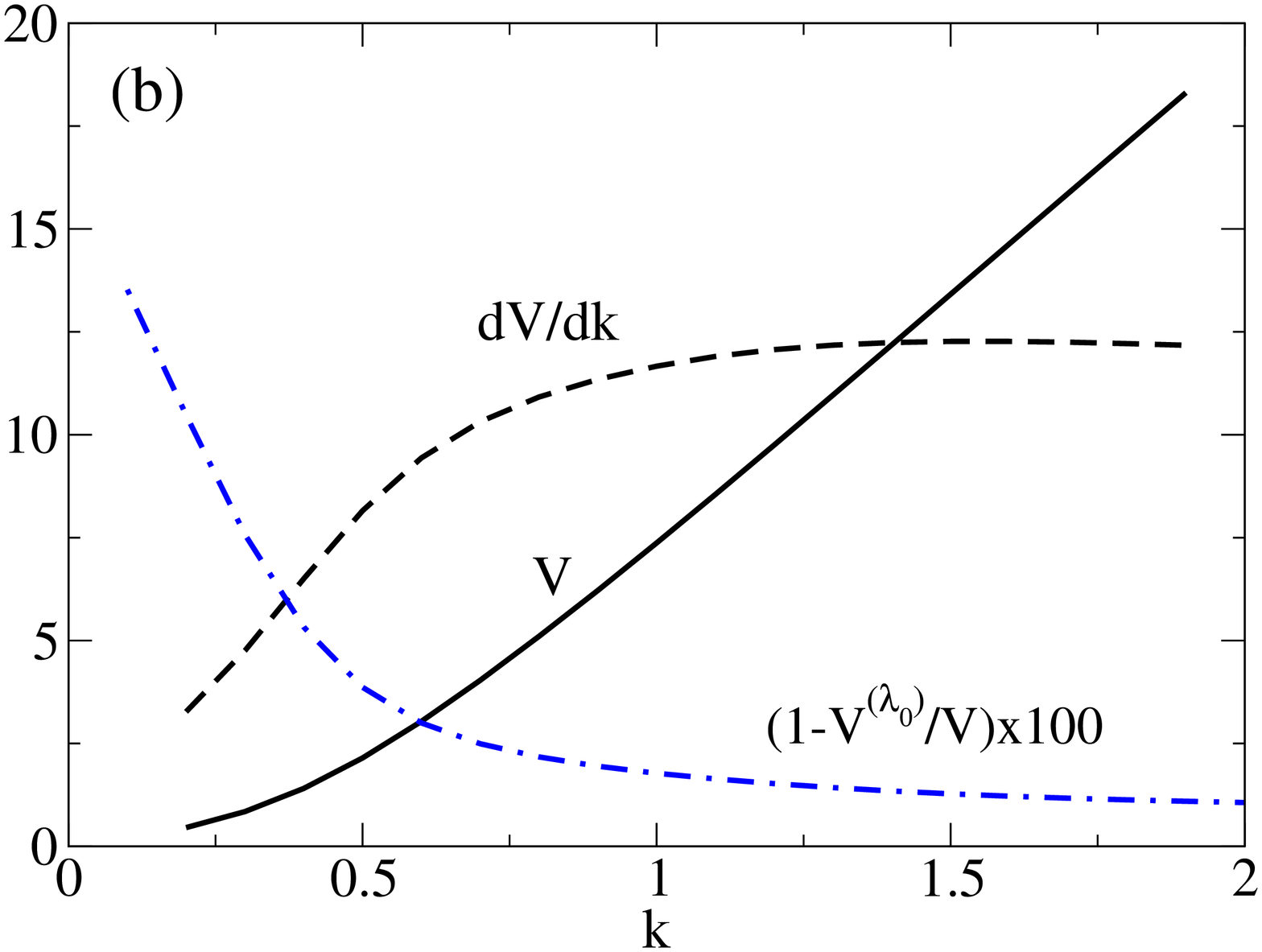,width=3in,height=3.5in,angle=-90} \\
\end{tabular}
\caption{BCS and PBCS results for $^{110}$Sn. (a) The largest ( $\lambda_0$) and the second largest ( $\lambda_1$) eigenvalues 
 of two-body density matrix as a function of the scaling factor $k$. (b) The average of the pairing force, $V$,
 its derivative to $k$ and the quantity $(1-V^{(\lambda_0)}/V) \times 100$ calculated in the BCS approximation.}
\end{figure}

In Fig. 2b are shown the BCS and PBCS results for the pairing energies and their derivatives to the pairing strength.
The pairing energy is defined as the average of the pairing interaction from which it is subtracted the self-energy
contribution. One observes that the pairing energies have a fast increase in the same region as for the largest eigenvalue, 
supporting the association of the latter to the evolution towards a paired phase.
In Fig. 2d it is also shown the dependence of the BCS pairing gap on the pairing strength. At variance with the largest
eigenvalue and the pairing energy, this quantity has a rather constant increase with the pairing strength.
 
In what follows we discuss  the properties of the two-body density matrix  taking as example a realistic
system, i.e. the nucleus $^{110}$Sn.  To  describe the pairing correlations in this nucleus we use as 
input for the Hamiltonian (1) the single-particle energies and the two-body pairing 
 matrix elements from Ref. \cite{volya}. In order to study how the strength of the interaction affects
 the eigenvalues of two-body density matrix, the  pairing matrix elements are scaled by a factor $k$ around
 the optimal physical value (corresponding to  $k$=1).  The results for the largest and the second largest 
 eigenvalues provided by BCS and PBCS approximations are displayed in Fig. 3a. The largest eigenvalue has its
 fastest  increase below k=1 and above the region where BCS has a nontrivial solution. As in the picket
 fence model discussed above, the largest eigenvalue and its derivative provided by PBCS evolve smoothly 
 across the region where the BCS solution sets in. This smooth behaviour of the largest eigenvalue is in keeping 
 with the fact that in finite system there is a gradual change of the structure of the ground state,
 from a HF-like state to a paired state, in contrast with the sudden phase transition predicted by BCS. 

\begin{figure}[ht]
\centering
\epsfig{file=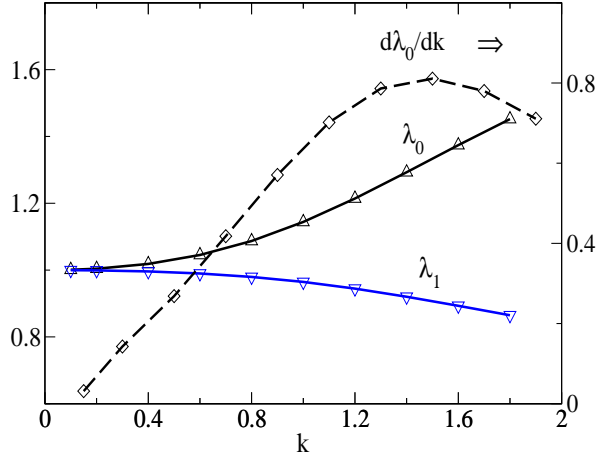,width=3in,height=3.5in,angle=-90} 
\caption{The exact results for the largest  and the
second largest eigenvalues  of two-body density matrix 
for $^{24}$O . The calculations are done with the  
matrix elements ( $J=0,T=1$) of the USDB interaction which
are scaled by the factor $k$. }
\end{figure}

 In Fig. 3b it is shown the average of the pairing force, $V$, and the 
 quantity $(1-V^{(\lambda_0)}/V) \times 100$. As in the case of Fig. 2b, one can observe that: 
(i) $ V $ and
 the largest  eigenvalue $\lambda_0$ have the fastest increase in the same region of the pairing strength;
 (ii) the largest contribution to $ V $ is coming from the largest eigenvalue.

In order to illustrate what happens in systems in which the like-particle pairing correlations
are weak, in Fig. 4 are shown the largest two eigenvalues for $^{24}$O. The results corresponds
to the exact diagonalisation performed for a pairing force extracted from the (J=0,T=1) part of the
realistic USDB interaction \cite{usd} which is scaled by a factor $k$. From Fig. 4 one can observe 
that at $k=1$ the ratio between the largest and the second largest eigenvalue are much smaller 
than in the case of $^{110}$Sn. In addition, the derivative of the largest eigenvalue indicates that 
in $^{24}$O the transition region to the paired phase is extended beyond  $k=1$. Apart from these
differences, related to the intensity of pairing correlations, these results shows the same pattern
as in the systems analysed above.

\section{Eigenvalues of density matrix for proton-neutron pairing interactions}

In atomic nuclei one can have pairing correlations not only between like-particles but also 
between protons and neutrons. These correlations are expected
to be the most relevant ones in nuclei with the same number of neutrons and protons since for these nuclei
the overlap between the neutron and proton wave functions is maximum. In these nuclei one commonly
considers two types of proton-neutron (pn) pairs, i.e. those with angular momentum $J=0$ and  isospin $T=1$
and those with $J=1$ and $T=0$ \cite{frauendorf,sagawa}. These pairs are called isovector and
isoscalar pn pairs, respectively.

The most general spherically symmetric isovector plus isoscalar pairing Hamiltonian reads as
\begin{eqnarray}
H=\sum_i\epsilon_i N_i&&+ 
\sum_{i,k} V^{01}(i,k) 
[\mathcal{A}^{+ 01}(i,i) \mathcal{\widetilde{A}}^{01}(k,k)]^{(0,0)}\nonumber\\
&&+ \sum_{i\leq i',k\leq k'} V^{10}(ii';kk') 
[\mathcal{A}^{+ 10}(i,i') \mathcal{\widetilde{A}}^{10}(k,k')]^{(0,0)}
\end{eqnarray}
and it refers to a system of protons and neutrons distributed over a set of orbitals 
$i=\{n_i,l_i,j_i\}$, where the standard notation for spherical single-particle states has been adopted. In this expression, $\epsilon_i$ and $N_i$ are the energy and the total particle number operator relative to the orbital $i$, respectively. Pair creation operators are defined as
\begin{equation}
\mathcal{A}^{+ JT}_{J_z,T_z}(i,i')= \frac{1}{\sqrt{(1+\delta(i,i'))}} [ a^+_i a^+_{i'} ]^{J,T}_{J_z,T_z},
\end{equation}
with $J, T$ being the angular momentum and isospin of the pair and $J_z,T_z$ their relative projections. For the pair annihilation operators we have adopted the standard definition ${\widetilde{A}}^{JT}_{J_z,T_z}(i,i')=(-1)^{J-J_z+T-T_z}{A}^{JT}_{-J_z,-T_z}(i,i')$.
The second and the third terms of Eq. (30) describe the isovector and the isoscalar pairing interactions, respectively. The Coulomb interaction between the protons has been neglected.

The Hamiltonian (30) is commonly used to investigate in $N=Z$ nuclei the presence of correlations of superfluid type
associated with isovector and isoscalar proton-neutron pairs (\cite{goodman_adv,goodman_prc,gerzelis}.
The issue we study here is whether for these pairs the density matrix has properties similar to those observed
for the like-particle pairing. In order to perform this study we construct the two-body density matrix in term of
pairs of arbitrary $(J,T)$ defined by
\beq
\rho^{(2)}_{JT}(ii',kk')= 
\sqrt{2J+1}\sqrt{2T+1}\langle \Phi^{(N)}_0 | 
[\mathcal{A}^{+ JT} (i,i') \mathcal{\widetilde{A}}^{+ JT}(k,k')]^{J=0,T=0} | \Phi^{(N)}_0 \rangle 
\eeq
with $i\leq i'$ and $k\leq k'$. 
Of interest for this study are the density matrices for $(J=0,T=1)$ and $(J=1,T=0)$.

As an illustration, we present in the following the results for the $N=Z$ nucleus $^{32}$S. 
As input for the Hamiltonian (30) we have used the single-particle energies and the matrix elements
for $(J=0,T=1)$ and $(J=1,T=0)$ extracted from  the  interaction USDB \cite{usd}.
 The density matrix has been calculated in correspondence with the exact
eigenstates of the Hamiltonian (30).

\begin{figure}[ht]
\centering
\begin{tabular}{cc}
\epsfig{file=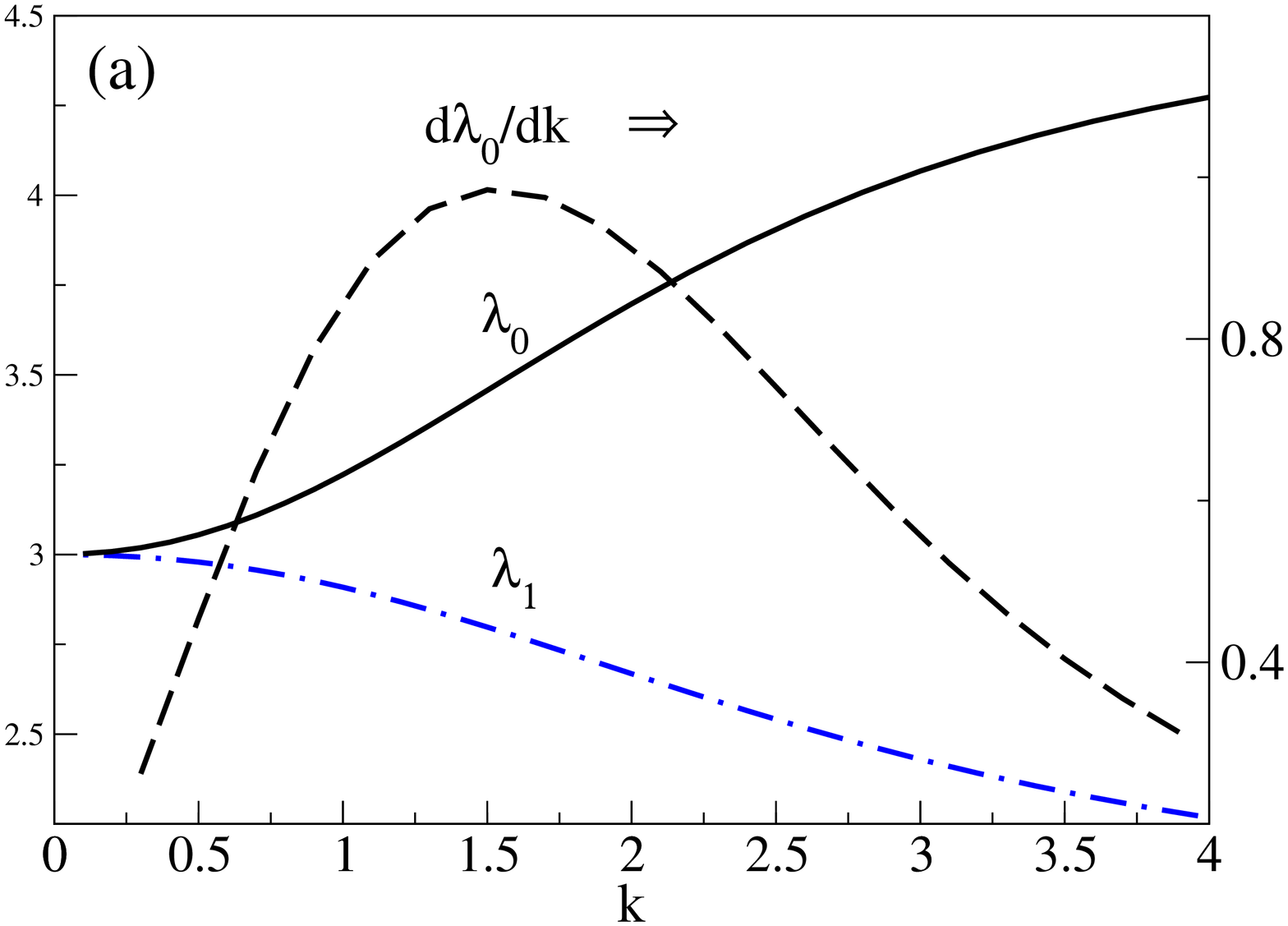,width=2.5in,height=3.5in,angle=-90} &
\epsfig{file=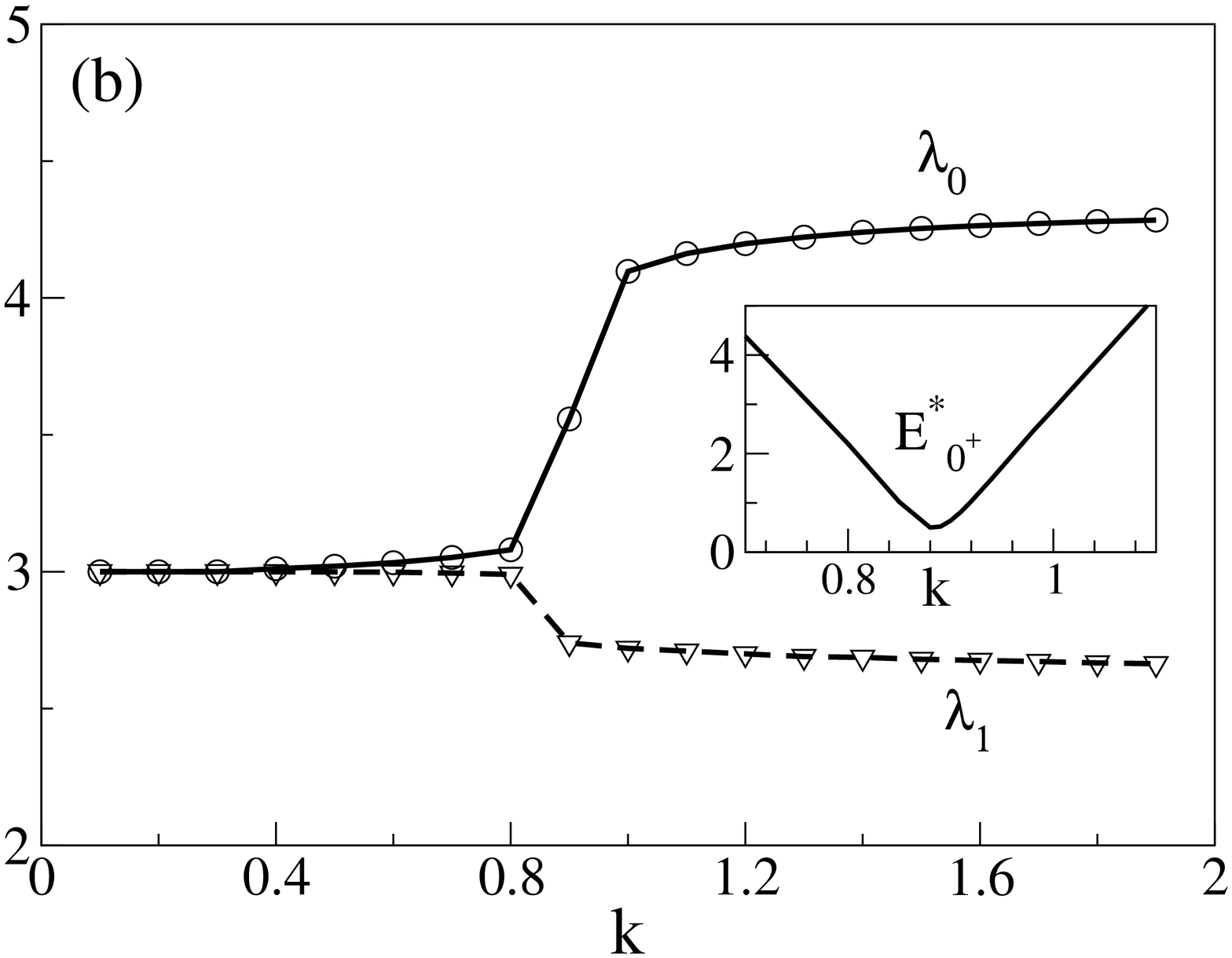,width=2.5in,height=3.5in,angle=-90} \\
\end{tabular}
\end{figure}
\begin{figure}[ht]
\centering
\begin{tabular}{cc}
\epsfig{file=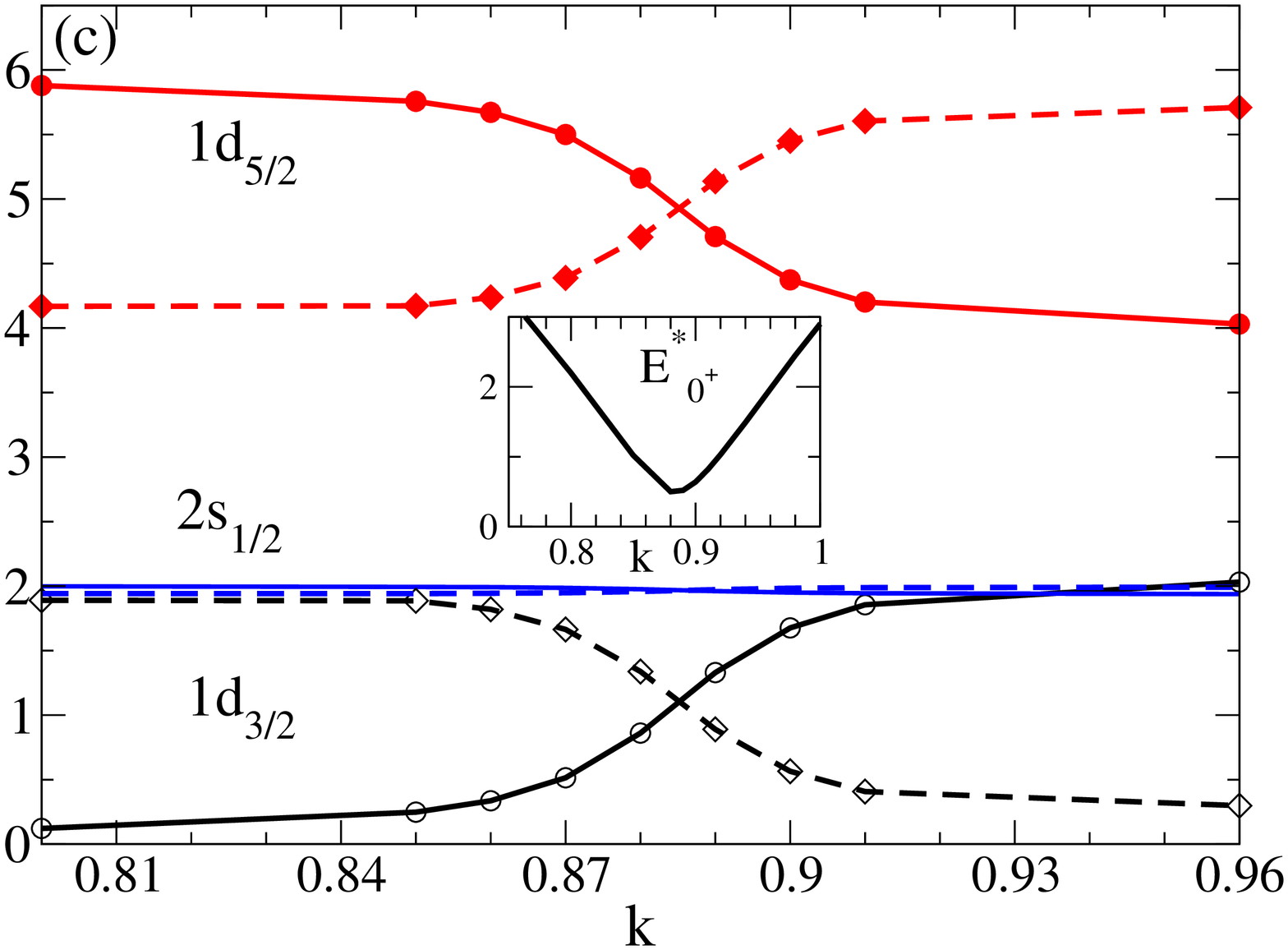,width=2.5in,height=3.5in,angle=-90} &
\epsfig{file=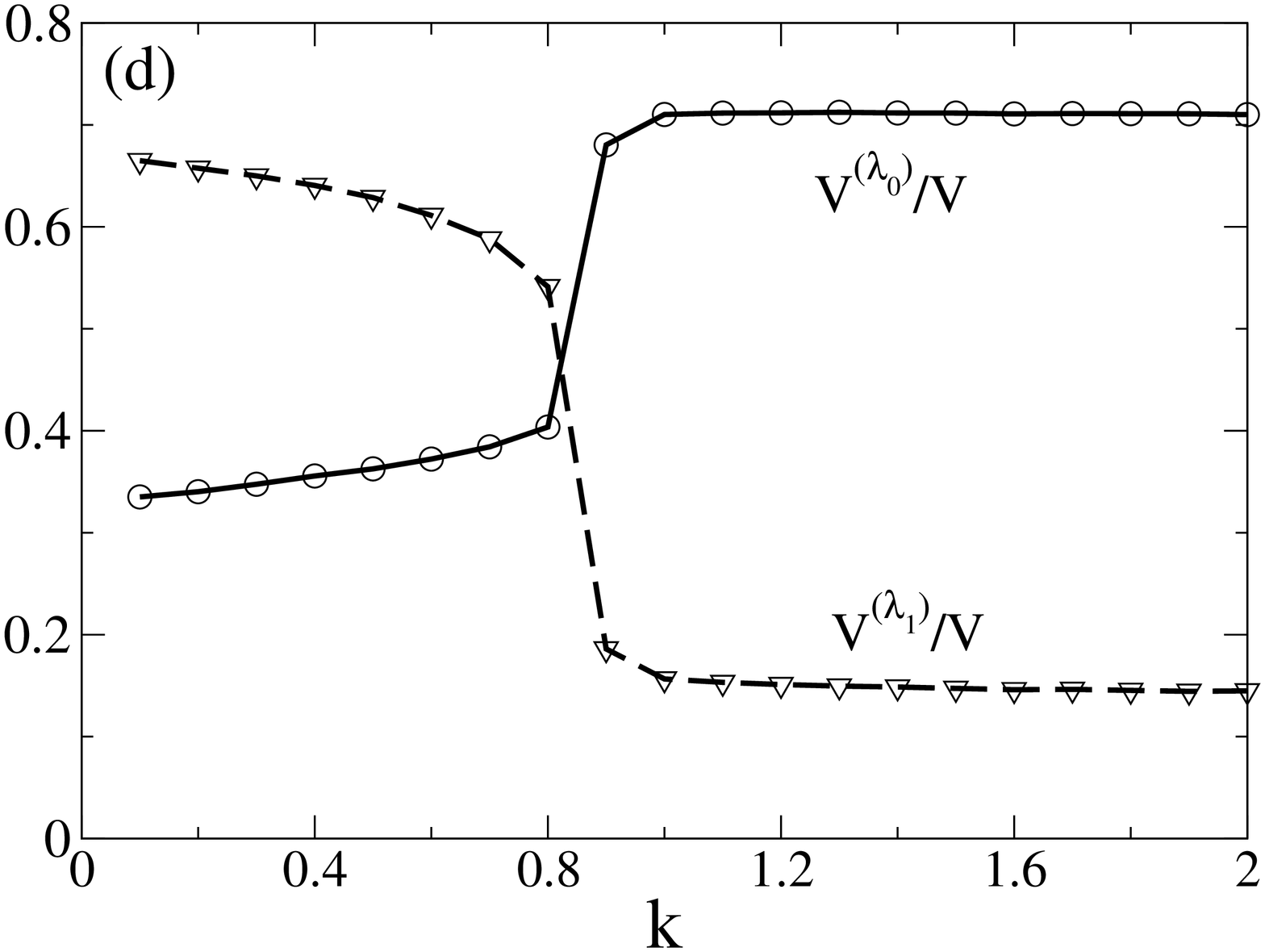,width=2.5in,height=3.5in,angle=-90} \\
\end{tabular}
\caption{
(a) The largest and the second largest eigenvalues of the two-body density matrix for
the isovector  pairing interaction scaled by the factor $k$. (b) The same as in (a) but for
the isoscalar pairing interaction. (c) The occupancy of the single-particle states of the
$sd$ valence shell in the case of a pure $T=0,J=1$ force. The insets in (b) and (c)
show the excitation energy of the first excited state $0^+$.
(d) The ratios $V^{(\lambda_0)}/V$ and $V^{(\lambda_1)}/V$ for the
isoscalar interaction. All results corresponds to $^{32}$S.}
\end{figure}

We begin by analysing the eigenvalues of the two-body density matrix  in the case of an
isovector interaction only. The strength of the interaction has been varied by multiplying all the $(J=0,T=1)$ USDB matrix
elements by a scaling factor $k$. The dependence of the largest two eigenvalues of
the density matrix on the scaling factor $k$ is shown in Fig. 5a. In the density matrix (32)
are added together the contribution of the proton-neutron, neutron-neutron and proton-proton
pairs which, due to the isospin invariance of the interaction, are equal to each other.
As a consequence, one observes that in the non-interacting limit the eigenvalues of the 
density matrix for $(J=0,T=1)$ are equal to 3. Fig. 5a shows that, with increasing the
interaction, the two largest eigenvalues behave as in the case of the nucleus 
$^{24}$O (Fig. 4), which has the same number of neutrons as $^{32}$S. Thus,
as expected from the isospin invariance, the properties of the density matrix for 
the isovector proton-neutron pairing are similar to those observed in the previous
section in the case of the neutron-neutron pairing. 

As a next step we analyse the results for the isoscalar interaction only, i.e. when from the USDB 
interaction we retain only the matrix elements with $(J=1,T=0)$. The results for the largest
two eigenvalues of the density matrix, as a function of the scaling factor $k$, are shown
in Fig. 5b. As in the case of isovector pairing one can see that in the non-interacting
case the eigenvalues are equal to 3. This is due to the fact that in the density matrix for
$(J=1,T=0)$ we are summing the three equal contributions of the pairs with $J_z=\{-1,0,1 \}$.
From Fig. 5b one can observe that up to $k \approx 0.8$ the largest two eigenvalues change
little compared to the non-interacting value. Then, between $k \approx 0.8$ and $k \approx 1.0$
the largest eigenvalue increases suddenly while the second largest decreases. 
Afterwards, for $k>1$, the two eigenvalues change  slowly with increasing the
interaction strength. Interestingly, in the region where the eigenvalues have a sudden 
change there is  a fast decrease of the energy of the first excited $0^+$ state. 
As seen in the inset of Fig. 5b, for $k \approx 0.88$ the energy of the excited state 
comes very close to the ground state energy, indicating a level crossing. 
This crossing is also reflected in the interchange between the occupancies of the spin-orbit partners $1d_{5/2}$ and $1d_{3/2}$ which is illustrated in Fig. 5c.

In Fig. 5d we show the quantities $V^{(\lambda_0)}$ and 
$V^{(\lambda_1)} $ relative to $ V $.
 We can see that the contribution 
of the largest eigenvalue to the average interaction is much smaller than for the
like-particle pairing. This is due to the fact that the amplitudes $f^{(0)}_i$ have mixed signs, 
so they do not add coherently in $V^{(\lambda_0)}$. This is also the case
for the other eigenvectors. Surprisingly, below $k \approx 0.8$ the
contribution of the largest eigenvalue is in fact much smaller than the contribution of 
the second largest eigenvalue. Since in this region the two eigenvalues are very close
to each other, this  means that the contribution arising from the $f^{(1)}_i$'s is larger than that from the $f^{(0)}_i$'s.
In Fig. 5d one can also notice that above $k \approx 0.8$ the two ratios depends very
weakly on the pairing strength. This is another indication that  the sudden  increase 
of the largest eigenvalue of the density matrix with the strength of
the isoscalar interaction is not related to a transition towards a paired phase.

\begin{figure}[ht]
\centering
\epsfig{file=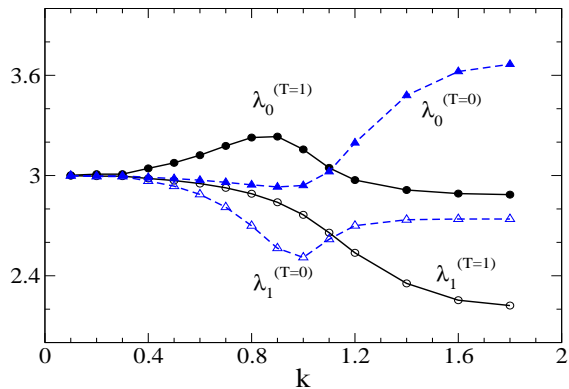,width=2.5in,height=3.5in,angle=-90}
\caption{
The largest two eigenvalues of the two-body density matrix for T=0 and T=1 pairs
obtained with the isovector plus the isoscalar pairing interactions scaled by 
the factor $k$. The results correspond to $^{32}$S. }
\end{figure}

In what follows we analyse the eigenvalues of the density matrix when the Hamiltonian (30) contains both the isovector and the isoscalar components. As in the calculations presented
above, for the two interactions we have considered the matrix elements with $(J=0,T=1)$ and $(J=1,T=0)$
extracted from the USDB interaction. Thus the relative strength of the interaction in the two channels
is fixed by the realistic USDB force. The calculations for the density matrix  are done by
scaling both interactions with the same factor $k$. The results for the largest and the second
largest eigenvalues of the density matrix, corresponding to the exact ground state, are shown in
Fig. 6.  One observes  that for $J=0$ the largest eigenvalue slowly increases
up to around $k = 0.8$ and then, for $k > 0.9$, it decreases towards the non-interacting value.
This decrease is not seen in the case of pure isovector interaction shown in Fig. 5a, which means
that this is an effect caused by the isoscalar pairing interaction .
On the other hand, as seen in Fig. 6,  the largest eigenvalue for $J=1$ has a shape rather
similar to the case of the pure isoscalar interaction. At a closer inspection one can notice, 
however, that the presence of the isovector force has some effects on the behavior of the
largest eigenvalue with $J=1$. One can indeed see that this slowly 
decreases  for $k < 1$, which is not so for the pure isoscalar force.
As it will be shown below, this decrease becomes significantly large in the presence of
the other multipoles of the two-body interaction. 

\begin{figure}[ht]
\centering
\epsfig{file=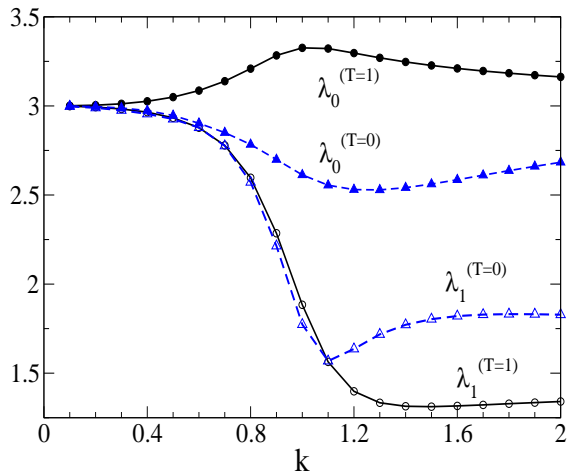,width=3in,height=3.5in,angle=-90} \\
\caption{The largest two eigenvalues of the two-body density matrix for T=0 and T=1 pairs
obtained with the USD interaction scaled by the factor $k$. The results corresponds
to the exact ground state of $^{32}$S.}  
\end{figure}

Finally we analyse the eigenvalues of the two-body density matrix for the $J=0$ and
$J=1$ pairs in the case of the full shell model Hamiltonian
\begin{equation}
H=\sum_i  \epsilon_i N_i  +  \sum_{i\leq i',k\leq k',J,T} V^{JT}(ii';kk')
\sqrt{2J+1}\sqrt{2T+1} 
[\mathcal{A}^{+ JT}(i,i') \mathcal{\widetilde{A}}^{JT}(k,k')]^{(0,0)},
\end{equation}
where the sum over $J,T$ includes all possible pairs.
The question of interest here is how the two-body density matrices for 
the ($J=0,T=1)$ and $(J=1,T=0)$ pairs are affected when, in addition to the interaction
in these two channels, one considers also all the other components of the shell model interaction. 
In order to study this issue, we have evaluated the two-body density matrices using the 
exact wave functions of the shell model (SM) Hamiltonian (33). 
In the SM calculations, performed with the code BIGSTICK \cite{bigstick}, we have used
the USDB interaction scaled by a factor $k$. The results for the largest and the second
largest eigenvalues of the density matrix in the channels $(J=0,T=1)$ and $(J=1,T=0)$ are
shown in Fig. 7. For $J=0$ the behavior of the largest eigenvalue is rather similar
to the one shown in Fig. 6, with the difference that for $k > 1$ the decrease is slower.
The situation is completely different for the $J=1$ pairing. In this case the largest eigenvalue
decreases very fast up to around $k=1.1$ and then it only slightly increases. This
shows that the peculiar fast increase of the largest eigenvalue observed in the case 
of the isoscalar and the isoscalar-isovector pairing interactions is washed out by the
other components of the SM interaction. 

\section{Summary and Conclusions}

In this paper we have presented a  detailed study  of the two-body 
density matrix corresponding to the interactions commonly
employed to describe pairing correlations in atomic nuclei.
After a general introduction
on the properties of the density matrix we have first analysed the eigenvalues of 
the two-body density matrix for various like-particle pairing interactions.
In all the cases investigated we have observed that the behaviour of 
the largest eigenvalue of the density matrix provides a clear indication of the 
transition from a normal to a paired phase. Indeed, at variance with the remaining 
eigenvalues, the largest eigenvalue has a fast and continuous increase in the
region of the pairing strength where a standard BCS approach predicts strong pairing correlations.
However, differently from  what happens in infinite systems, the largest 
eigenvalue is not found too much different in magnitude as compared to the other eigenvalues, 
especially in the  weak coupling regime. In spite of that, the largest eigenvalue
is responsible for the dominant contribution in the expectation value of the pairing interaction, as in
the case of infinite systems. This is due to the fact that the eigenfunction corresponding
to the largest eigenvalue has a coherent structure, i.e. all its components have the same
sign. In the like-particle pairing case we have also noticed a close agreement between 
exact and PBCS results for the largest eigenvalue of the density matrix. This finding 
is consistent with the good agreement found for the correlation energies in previous 
studies. It is worth mentioning that a close agreement between PBCS and the exact results
can be obtained only when the pairing acts in a limited window around the Fermi level
of the order of the pairing gap \cite{sandulescu_bertsch}. 

In the second part of the paper we have studied the properties of the density matrix for
the isoscalar ($J=0$, $T=1$) and the isovector ($J=1$, $T=0$) proton-neutron pairing interactions.
For this study we have considered as example the nucleus $^{32}$S.  
The results for the isovector interaction are rather similar to those for the 
like-particle pairing. Very different results have been obtained instead  for the
isoscalar interaction. In this case the largest $J=1$ eigenvalue has a sudden increase when the strength of the interaction is scaled around the
physical value. A detailed analysis shows, however, that this sudden increase of the largest 
eigenvalue is in fact related to the crossing between the ground state and the first 
excited $0^+$ state and not to a transition towards a paired phase. 

We have also analysed the eigenvalues of the density matrix when the isovector and the isoscalar
pairing interactions are acting together. The results for this case are not significantly
different from those corresponding to the pure pairing forces mentioned above, except for the decrease of the
largest $J=0$ eigenvalue for large values of the pairing strength.

Finally, in the last part of the paper we have studied the properties of the two-body density matrix  
for a general interaction. As example we have taken the standard shell-model two-body
interaction which is commonly applied to describe the nucleus $^{32}$S. We found that the largest $J=1$ 
eigenvalue is decreasing very fast with the increase of the interaction strength, arriving to values 
below those at the non-interacting regime for strengths around the physical value. 
This is an indication that no long-range two-body correlations can be associated with $J=1$ pairs in the
case of the general interaction . 

In the present study we have focused on the two-body correlations associate to proton-neutron pairing interactions.
However, in the N=Z systems these interactions generate also 4-body correlations \cite{soloviev,flower,
bremond,eichler,dobes_pittel,chapman,qcm_nez,qcm_plb,qm_t1}. In particular, as shown recently, the isovector plus isoscalar Hamiltonian (30) can be treated accurately in 
terms of alpha-like quartets built by two protons and two neutrons coupled to total isospin $T=0$, rather than
in terms of Cooper pairs \cite{qcm_t0t1}. It has also been shown that $T=0$ alpha-like quartets are the main
building blocks  for systems governed by the general shell model  Hamiltonian (33) \cite{hasegawa1,qm_prl,qcm_epja}. 
These studies clearly show that  4-body correlations play a key 
role in the N=Z nuclei. How these correlations are reflecting in the 4-body density matrix is an interesting
issue which will be addressed in a future publication.

\vskip 0.4cm
\noindent
{\bf Acknowledgements}
\vskip 0.2cm
\noindent
N. S. would like to thank for the hospitality of Royal Institute of Technology, Stockholm, 
where this paper was mainly written. 
This work was supported by the Romanian National Authority for Scientific Research, CNCS  UEFISCDI, 
Project Number PN-II-ID-PCE-2011-3-0596.


\begin{thebibliography}{10}
\bibitem{penrose}  O. Penrose, Phil. Mag. {\bf 42}, 1373 (1951).
\bibitem{penrose_onsager} O. Penrose and L. Onsager, Phys. Rev. {\bf 104}, 576 (1956).
\bibitem{yang} C. N. Yang, Rev. Mod. Phys. {\bf 34}, 694 (1962).
\bibitem{bcs} J. Bardeen, L. N. Cooper, and J. R. Schrieffer, Phys. Rev. {\bf 108}, 1175(1957).
\bibitem{bohr_mottelson} A. Bohr and B. Mottelson, Nuclear Structure (World Scientific, 1998).
\bibitem{richardson} R. W. Richardson and N. Sherman, Nucl. Phys. {\bf 52}, 221 (1964).
\bibitem{bayman}B. F. Bayman, Nucl. Phys. {\bf 15}, 33 (1960).
\bibitem{pbcs} K. Dietrich, H. J. Mang, and J. H. Pradal, Phys. Rev. {\bf 135}, B22 (1964).
\bibitem{langanke} K. Langanke, D. J. Dean, P. B. Radha, S. E. Koonin, Nucl. Phys. A {\bf 602}, 244 (1996). 
\bibitem{frauendorf}
S. Frauendorf and A. O. Macchiavelli, Prog. Part. Nucl. Phys. {\bf 78}, 24 (2014).
\bibitem{sagawa} H. Sagawa, C. L. Bai, and G. Colo, Phys. Scripta {\bf 91}, 083011 (2016).
\bibitem{leggett} A. Leggett, Quantum Liquids (Oxford University Press, 2006).
\bibitem{ginzburg_landau}
V.L. Ginzburg and L.D. Landau, Zh. Eksp. Teor. Fiz. {\bf 20}, 1064 (1950).
\bibitem{ns_kappa0} N. Sandulescu, P. Schuck, X. Vinas, Phys. Rev. C {\bf 71}, 054303 (2005).
\bibitem{ns_kappa} N. Pillet, N. Sandulescu, P. Schuck, Phys. Rev. C {\bf 76}, 024310 (2007).
\bibitem{schechter} M. Schechter, Y. Imry, Y. Levinson, and J. von Delft, Phys. Rev.
B {\bf 63}, 214518 (2001).
\bibitem{delft} J. von Delft and D. C. Ralph, Phys. Rep. {\bf 345}, 61 (2001).
\bibitem{volya} A. Volya, V. Zelevinsky, Phys. Lett. B {\bf 574}, 27 (2003).
\bibitem{sandulescu_errea} N. Sandulescu, B. Errea, J. Dukelsky, Phys. Rev. C {\bf 80}, 044335 (2009).
\bibitem{usd} B.A. Brown and W.A. Richter, Phys. Rev. C {\bf 74}, 034315 (2006).
\bibitem{goodman_adv} A. L. Goodman, Adv. Nucl. Phys. {\bf 11}, 263 (1979).
\bibitem{goodman_prc} A.L. Goodman, Phys. Rev. C {\bf 60}, 014311 (1999).
\bibitem{gerzelis} A. Gezerlis, G.F. Bertsch, Y.L. Luo, Phys. Rev. Lett. {\bf 106}, 252502 (2011).
\bibitem{bigstick} C. W. Johnson,W. E. Ormand, K. S. McElvain, and H. Z. Shan, arXiv: 1801.08432.
\bibitem{sandulescu_bertsch} N. Sandulescu, G. Bertsch, Phys. Rev. C {\bf78}, 064318 (2008).
\bibitem{soloviev} V.G. Soloviev, Nucl. Phys. {\bf18}, 161 (1960).
\bibitem{flower} B. H. Flowers and M. Vujicic, Nucl. Phys. {\bf49}, 586 (1963).
\bibitem{bremond} B. Bremond, J.G. Valatin, Nucl. Phys. {\bf41}, 640 (1963).
\bibitem{eichler} J. Eichler and M. Yamamura, Nucl. Phys. A {\bf182}, 33 (1972).
\bibitem{dobes_pittel} J. Dobes and S. Pittel, Phys. Rev. C {\bf57}, 688 (1998).
\bibitem{chapman} R.R. Chasman, Phys. Lett. B {\bf577}, 47 (2003).
\bibitem{qcm_nez} N. Sandulescu, D. Negrea, J. Dukelsky, C. W. Johnson, 
Phys. Rev. C {\bf85}, 061303(R) (2012).
 \bibitem{qcm_plb} N. Sandulescu, D. Negrea, and D. Gambacurta, Phys. Lett. B {\bf751}, 348 (2015).
\bibitem{qm_t1} M. Sambataro and N. Sandulescu, Phys. Rev. C {\bf 88}, 061303(R) (2013).
\bibitem{qcm_t0t1} M. Sambataro, N. Sandulescu, C.W. Johnson, Phys. Lett. B{\bf740}, 137 (2015).
\bibitem{hasegawa1} M. Hasegawa, S. Tazaki, and R. Okamoto, Nucl. Phys. A {\bf592},
45 (1995).
\bibitem{qm_prl} M. Sambataro and N. Sandulescu, Phys. Rev. Lett. {\bf 115}, 112501 (2015).
\bibitem{qcm_epja} M. Sambataro and N. Sandulescu, Eur. Phys. J. A {\bf 53}, 47 (2017).
\end{thebibliography}
\end{document}